\documentstyle[floats, psfig, prl, aps]{revtex} 
\begin{document} 
\draft 
\twocolumn[\hsize\textwidth\columnwidth\hsize\csname 
@twocolumnfalse\endcsname 
 
\title{Universality in the Cross-over between Edge Channel 
and Bulk Transport  
in the Quantum Hall Regime} 
 
\author{J. Oswald, G.Span, F.Kuchar} 
\address{Institute of Physics, University of Leoben, Franz 
Josef Str. 18, A-8700 Leoben, Austria} 
 
\date{\today} 
 
\maketitle 
 
\begin{abstract} 
We present a new theoretical approach
for the integer quantum Hall 
effect, which is able to describe
the inter-plateau transitions as well as the transition to the Hall %%@
insulator. We find 
two regimes (metallic and insulator like) of the top Landau level, in which %%@
the 
dissipative bulk current appears in different directions. The 
regimes are separated by a temperature invariant point.

\end{abstract} 
 
\pacs{73.50.-h,73.40.Hm,72.10.-d,71.30.+h} 
\vskip2pc 
 
] 
 
Even more than 15 years 
after the discovery of the integer quantum Hall effect (IQHE) 
in two-dimensional electronic systems \cite{Klitz}, the nature 
of the transitions between 
adjacent QH plateaus is still a controversial question. 
Already years ago Kivelson et al \cite{6} developed a 
global phase diagram, which maps out insulator and 
QH-liquid phases onto a magnetic field-disorder plane. 
However, a quantitative modeling of the complete transport 
regime of the 
IQHE has not been given so far. While the quantized values of 
the Hall resistance are well described by the 
edge channel (EC) picture \cite{1}, it is widely believed that 
the EC-picture is insufficient to describe also the transport 
regime between the IQHE plateaus. The ongoing studies of 
the inter-plateau transitions are particularly stimulated by 
recent results of the work on the Hall insulator 
(HI)\cite{2,3,4}. Using QHE samples with not 
too low disorder, the HI regime is entered directly from the 
$\nu=1$ IQHE regime without observing 
the fractional QHE. An analysis of the transport ranging 
from the HI to the adjacent QH liquid regime suggests the 
existence of a close relation between the transport 
behavior in the two regimes \cite{4}:  By defining a critical  
filling factor $\nu_c$ it is possible to distinguish two 
regimes, that are coupled by the relation $\rho_{xx}(\Delta 
\nu)=1/\rho_{xx}(-\Delta \nu)$, where $\Delta \nu$ is the 
filling factor relative to $\nu_c$. Another 
important experimental fact is the existence of a critical 
longitudinal resistivity $\rho_{xx}^c$, which appears at the 
transition point from the QH-liquid to the HI regime \cite{2}. 
This critical point $\nu_c$ is indicated by the crossing of the 
temperature dependent $\rho_{xx}$ traces and the value 
of $\rho_{xx}^c$ was found to be close to $h/e^2$. Using a 
tensor based analysis of 
the experimental data, Shahar et al. \cite{5} have been 
able to extract the contribution of the top LL (referred 
to as $\rho_{xx}^{top}$) to the total $\rho_{xx}$ in the 
transition regime between the $1^{st}$ and $2^{nd}$ QH-
plateau. They found that $\rho_{xx}^{top}$ 
shows the same behavior like 
$\rho_{xx}^{ins}$  in the HI regime, namely a monotonous 
increase with increasing magnetic field without any peak-like 
behavior. It was possible 
to collapse all temperature dependent traces onto each 
other by plotting $\rho_{xx}^{ins}$  as well as 
$\rho_{xx}^{top}$ 
with respect to $(\nu - \nu_c)T^{-\kappa}$ using the same  
$\kappa = 0.45$. Another experimental fact is that 
$\rho_{xy}$ 
remains quantized on the  $\nu=1$ plateau also in 
the HI regime below the critical filling factor, while 
$\rho_{xx}^{ins}$ already steeply rises. Furthermore Shahar 
et al 
demonstrated, that the conductivity components in the HI 
regime ( $\sigma_{xx}^{ins}$ and $\sigma_{xy}^{ins}$) as 
well as the extracted components for the top LL 
($\sigma_{xx}^{top}$ and $\sigma_{xy}^{top}$) for the 1 
$\rightarrow$ 2 plateau transition fulfill a semicircle 
relation ($\sigma_{xx}^2 + \sigma_{xy}^2 \propto 
\sigma_{xy} $) \cite{7}. 
 
	In this letter we show for the 
first time, that by 
a modification of the Landauer- B\"uttiker formalism \cite{1},
which is based on a new formulation of backscattering, 
it is possible to model the full IQHE behavior, i.e. the Hall
plateaus as well as the transport regime between them. Several
of the above referenced experimental facts can be modeled 
without assuming a particular function for the dependence of the 
backscattering on the filling of the Landau levels. Introducing
an exponential function for this dependence, further details of
the experimental findings are reproduced.  

Several attempts for 
modeling a four-terminal experiment with a discrete 
backscattering barrier or a disordered region between 
ideal conductors  have been made already \cite{1,10,Kivel}. As an 
example, for the case of a discrete backscattering barrier 
in a 4-terminal arrangement B\"uttiker\cite{1,10} obtains 
$R_{xx}=(h/e^2) \left[ R/(N \cdot T) \right]$, where $N$ is 
the number of channels, $R$ and $T$ are the reflection 
and transmission coefficients of the barrier. 
The Landauer- B\"uttiker formalism is completely general,
where the transmitting channels are not necessarily edge
channels. However,
for the explanation of the quantized Hall resistance values
B\"uttiker introduces the EC-picture.
We also use this picture \cite{ec1} as an input to our model 
in which we use an alternative representation of 
backscattering. Details of our approach can be obtained
from \cite{8}. The essence of  \cite{8} is
that the coupling of the states at opposite edges
via backscattering leads to a coupling between $R_{xx}$ and 
$R_{xy}$ according to $R_{xx}=P \cdot R_{xy}$, where 
$P$ is a novel backscattering parameter \cite{9}. 
For the case of a $\it{single}$ LL, which is
represented by a single pair of ECs, we substitute $R_{xy}$ by 
$h/e^2$: 
 
\begin{equation} 
R_{xx}=P\frac{h}{e^2} 
\label{Eq1} 
\end{equation} 
							
Eqn.\ref{Eq1} agrees perfectly with B\"uttikers above result:  
The factor $R/T$ can be identified with $P$. Since we have 
$N=1$ and $T= 1-R$, a variation of 
$R$ between 0 and 1 corresponds to a variation of $P$ 
between 
0 and $\infty$, which is the appropriate range for our model.  
 
	For a standard QH-system, as e.g. in 
AlGaAs/GaAs, backscattering appears only in the top 
LL in the regime between plateaus, while the transport in the 
lower LLs remains dissipation-less. For a transport model 
in the EC-picture one has therefore to combine one pair of 
ECs with non-zero backscattering ($P>0$) and a set of EC-
pairs without 
backscattering ($P=0$). $R_{xx}$ and $R_{xy}$ of the 
complete system must finally result from the current 
distribution between both EC systems \cite{11}. For 
treating these parallel systems we use the components of 
the conductance tensor ($G_{xx}$ and $G_{xy}$), which 
can be obtained from the components of the resistance 
tensor ($R_{xx}$ and $R_{xy}$) by the well known relation 
$G_{xx, yx}=R_{xx, xy} \cdot (R_{xx}^2+R_{xy}^2)^{-1}$. 
The use of this equation means that we restrict our 
analysis to the case of a symmetric behavior where 
$R_{yy}$ = $R_{xx}$. In comparison to classical transport 
this corresponds to the case of a quadratically shaped 
conductor. Consequently the equations are formally 
identical with the equations for the resistivities $\rho_{xx}$, 
$\rho_{xy}$ and conductivities $\sigma_{xx}$, 
$\sigma_{xy}$. By use of Eqn.\ref{Eq1} we get for the top LL 
 
\begin{equation} 
G_{xx}^{top}=\frac{e^2}{h} \cdot \frac{P}{1+P^2} 
\label{Eq2} 
\end{equation} 
Due to the absence of backscattering in the lower LLs we 
have $G_{xx}^{low}=0$ and therefore the total $G_{xx}$ is 
given by   $G_{xx}^{top}$. In an analogous way we 
calculate the Hall components  
 
\begin{mathletters} 
\begin{equation} 
G_{xy}^{top} = \frac{e^2}{h} \cdot \frac{1}{1+P^2} 
\label{Eq3a}  
\end{equation} 
\begin{equation} 
G_{xy}^{low} = \frac{e^2}{h} \cdot \bar{\nu} 
\label{Eq3b} 
\end{equation} 
\end{mathletters} 
where $\bar{\nu}$ is the number of filled LLs below the top 
LL. The total Hall conductance $G_{xy}$ is given by the 
sum of Eqn.\ref{Eq3a} and Eqn.\ref{Eq3b}. Now, using 
$R_{xx,xy}=G_{xx,yx}\cdot (G_{xx}^2+G_{xy}^2)^{-1}$ 
we 
obtain: 
 
\begin{mathletters} 
\begin{equation} 
R_{xx}=\frac{h}{e^2} \cdot \frac{P}{(\bar{\nu} +1)^2 + 
(\bar{\nu} \cdot P)^2} 
\label{Eq5a} 
\end{equation} 
\begin{equation} 
R_{xy}=\frac{h}{e^2} \cdot \left[ \bar{\nu} + 
\frac{1}{1+P^2}\right]  \cdot \left[\bar{\nu}^2 + 
\frac{2\bar{\nu}+1}{1+P^2}\right] ^{-1} 
\label{Eq5b} 
\end{equation} 
\end{mathletters} 
where the backscattering parameter $P$ depends on the 
partial filling $\nu^{top}$ of the top LL.  Even without 
knowing yet the function $P(\nu^{top})$, one can directly 
see that Eqn.\ref{Eq2} and Eqn.\ref{Eq3a} fulfill the 
semicircle relation $G_{xx}^2+G_{xy}^2 \propto G_{xy}$, 
which is valid also for the 
complete system. It was 
experimentally found to be valid for the top LL 
as well as for the HI-regime\cite{5}. 
 
\begin{figure} 
\centerline{\psfig{figure=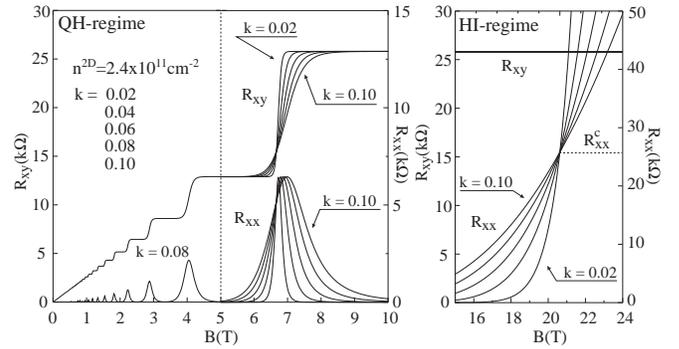,width=86mm}} 
\caption{$R_{xx}$ and $R_{xy}$ calculated according to 
Eqn.\ref{Eq5a},\ref{Eq5b} for a sheet carrier density of 
$n^{2D}=2.4x10^{11}cm^{-2}$ and different factors 
$k$ in the exponent of $P(\Delta \nu)$. The range below 
B=5T shows just the traces for k=0.08, the range above 
B=5T shows the traces for all different k values as given in 
the figure. The HI-regime is shown separately with a 
different $R_{xx}$ scale on the right.}  
\label{Fig1} 
\end{figure}

	It is widely accepted that in the plateau transition 
regions backscattering in the top LL enables bulk conduction if 
the Fermi level ($E_F$) is near the center of the 
broadened LL. In the bulk a transition to an 
insulating state occurs if  $E_F$ moves out of the 
center on either side \cite{12}. However, in a model which 
is based on the EC-picture, a symmetry in $E_F$ with respect 
to the 
center of the top LL 
cannot exist: By starting with $E_F$ above a completely filled 
top LL, ECs are formed and the transport can be described by 
the EC-picture without backscattering.  
With $E_F$ approaching the center 
of the LL, dissipation because of backscattering becomes 
possible and finally with 
$E_F$ moving below the center of the 
top-LL the associated pair of ECs disappears.  We will show that 
despite this asymmetric behavior of the top LL, the overall 
behavior of 
the total system can be still symmetric. Considering 
Eqn.\ref{Eq2} one can see that $G_{xx}$ is proportional to 
$P$ for $P \ll 1$, while it changes to a reciprocal 
dependence on $P$ for $P \gg 1$. For a symmetric 
$G_{xx}$ we have to look for a suitable monotonous 
function $P(\nu^{top})$ which is able to produce such a 
symmetric behavior around $\nu_c$. To get perfect symmetry, the 
form of Eqn.\ref{Eq2} requires a function which fulfills the 
relation $P(\Delta \nu)=1/P(-\Delta \nu)$ with $\Delta \nu = 
\nu ^{top}-\nu_c$. It is easily seen, 
that the only function which is also in agreement with the 
experimental observations \cite{4} is of the form 
\cite{wende}: 
 
\begin{equation} 
P(\Delta \nu)= \exp(-{\Delta \nu}/k) 
\label{Eq6} 
\end{equation} 
with k being a constant but possibly temperature
dependent factor. Since the maximum of 
$G_{xx}$  is identified with the 
center of the top LL, $\nu_c$ corresponds to half 
filling. From Fig.\ref{Fig1} one can see that the calculation 
based on Eqn.\ref{Eq5a}, \ref{Eq5b} and Eqn.\ref{Eq6}
reproduces 
very well the typical traces known from the experimental 
curves at different temperatures.

\begin{figure} 
\centerline{\psfig{figure=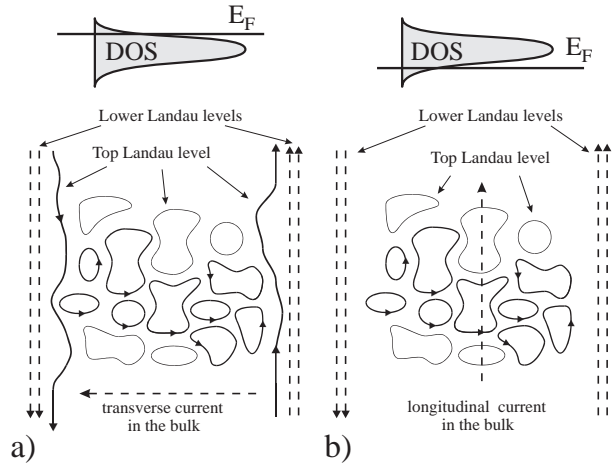,width=80mm}} 
\caption{a) Edge channel conduction in the top-LL in the 
presence of localized magnetic boundstates. The 
transport across the loops appears as a transverse 
current, which acts as a backscattering process. b) 
Conduction in the top-LL in the presence of localized 
magnetic boundstates but in absence of an associated 
EC. In contrast to the situation sketched in a),  the 
transport across the loops appears now as a longitudinal 
current. The ECs of the lower LLs are indicated by the 
dashed arrows. The relative position of the Fermi level 
with respect to the LL is indicated at the top of the figure.} 
\label{Fig2} 
\end{figure}

	Already without needing the particular function of
Eqn.\ref{Eq6} we get a number of important results.
We can consider two 
regimes which are divided by the point at which $P(\Delta \nu)=1$:
The regime $0<P<1$ corresponds to 
$E_F$ above the center of the top LL, while $P>1$ corresponds to 
 $E_F$ below the center of the top LL. 
Fig.\ref{Fig2}a and Fig.\ref{Fig2}b show schematically the 
situation in the two regimes:
While $E_F$ moving 
towards the center of the broadened top-LL 
(Fig.\ref{Fig2}a), localized magnetic bound states 
are created in the bulk region in addition to the 
associated pair of ECs. Therefore some 
transport across those loops by tunneling becomes possible, 
which finally enables backscattering. According to 
Eqn.\ref{Eq1}, $R_{xx}^{top}$ is directly proportional to the 
backscattering rate in this regime. For describing this type of 
transport in the bulk 
region, basically a network model such as e.g. that one of 
Chalker and Coddington \cite{13} would be suitable. A 
situation 
with $E_F$ below the LL center is 
schematically shown in Fig.\ref{Fig2}b with one major 
difference to Fig.\ref{Fig2}a, namely that the associated 
EC-pair is not present, while the transport mechanism in the
bulk itself may remain the same. 
Consequently the transport in 
the bulk does no longer act as a coupling between 
opposite edges, but  may contribute now via a current in the 
longitudinal direction instead. 
Characterizing the 
dissipative transport through the bulk by a conductivity 
$\sigma_{bulk}$, we get basically $R_{xx}^{top}\propto 
\sigma_{bulk}$ for $E_F$ above the LL center and 
$R_{xx}^{top}\propto \sigma_{bulk}^{-1}$ for $E_F$ below 
the LL center. Consequently any influence of an eventually 
existing temperature dependence of $\sigma_{bulk}$  on 
the longitudinal transport properties must appear with 
opposite sign in the two regimes. This implies 
that there must be a cross-over of the two 
regimes, where the temperature dependence of $R_{xx}$ 
is canceled.  In this way our model indicates correctly
the existence of metallic like and insulator like regimes.
One can also interpret 
the two regimes as two different phases of the top LL with 
perpendicular directions of the dissipative bulk current.
This is a striking agreement with Ruzin et al. \cite {7}, who 
also found, that for a correct description of the transport 
behavior the bulk current directions in both phases must be 
perpendicular to each other.
It is easily found that the critical point in the cross-over 
regime occurs at P = 1. According to Eqn.\ref{Eq1} 
this means that at the critical point $R_{xx}^{top}$ 
approaches the quantized value $h/e^2$. $P=1$ 
also means that for the transport in a single LL  
$G_{xx}=G_{xy}=0.5e^2/h$, in agreement with 
Ref.\cite{14}.  

	In \cite{5} also the $R_{xx}$ peak between the 
$1^{st}$ and $2^{nd}$ plateau has been analyzed. It has 
been found that the maximum value is 
$h/4e^2$ while $R_{xx}^c$ at the critical point appears is
 $h/5e^2$. In our model the critical point appears at $P 
= 1$, for which we get a value of $R_{xx}^c= h/5e^2$, in 
agreement with \cite{5}. Considering the maximum of 
Eqn.\ref{Eq5a} for $\bar{\nu}=1$, we find $P =2$, which 
leads to $R_{xx}^{max} = h/4e^2$ , also in agreement with 
\cite{5}.

Using the particular function $P(\Delta \nu)$ according  
to Eqn.\ref{Eq6}, we can
go a step further: With the help of Eqn.\ref{Eq1} we obtain 
$R_{xx}^{top} = (h/e^2) \cdot  \exp(-{\Delta \nu}/k)$, which 
is a monotonous function and covers both regimes $P>1$ and
$P<1$. 
Now we can also consider the principal 
behavior of $\sigma_{bulk}$ 
in the tails of the LL ($P \gg 1$ and $P \ll 1$) by using 
$\sigma_{bulk} \propto R_{xx}^{top}$ for $\Delta \nu > 0$ 
and $\sigma_{bulk} \propto 1/R_{xx}^{top}$ for $\Delta \nu 
< 0$. As generally expected for pure bulk transport,
we obtain a symmetric function around the LL-center 
$\sigma_{bulk}(\Delta \nu) \propto  \exp (-|\Delta \nu| /k)$. 
Thus it is demonstrated, that our model provides the
correct framework
 to include also dissipative bulk transport.

The experimental evidence for the non-symmetric 	
transport behavior of $R_{xx}^{top}$ comes with
$E_F$ in the lowest LL ($\bar{\nu}=0$, see Eqn.\ref{Eq5a}).
There $R_{xx}$ is identical to $R_{xx}^{top}$ and
increases monotonically with decreasing filling factor.
This is exactly  the regime of the HI, which has been 
experimentally very well investigated already: 
$R_{xx}^{ins}$ has been indeed found to be 
monotonously increasing without any peak behavior and 
$R_{xy}$ stays at the quantized value $h/e^2$ \cite{5}, in 
agreement with Eqn.\ref{Eq5b} for $\bar{\nu}=0$. 
Therefore we can interpret the behavior in the HI regime 
to be a direct consequence of the asymmetric transport 
behavior of a single LL. Since in our model the transition 
to the HI as well as the inter-plateau transitions are 
described by the same function $P(\Delta 
\nu)$, the experimentally observed equivalent behavior of 
$R_{xx}^{top}$ and $R_{xx}^{ins}$ \cite{5} is an inherent 
property of our model. 
 
The fact, that the temperature dependence 
disappears at a certain point, suggests that the 
temperature $T$ enters only the factor $k$ in the exponent 
of Eqn.\ref{Eq6}. Moreover, $\Delta \nu = 0$ in Eqn.\ref{Eq6} 
means that $P = 1$ and therefore $R_{xx}^c = h/e^2$ 
(Eqn.\ref{Eq5a} for $\bar{\nu} = 0$), in agreement with 
Ref.\cite{2}. This is also evident from Fig.\ref{Fig1}, where 
the traces cross each other at $R_{xx}^c = h/e^2$ (at 
$B=20T$). 
 
	A widely used basis for the discussion of experimental
data is the plot of the $\rho_{xx}$ peak width $\Delta B$
 as a function
of temperature. In this context we analyze the width of the 
$G_{xx}$ peak, which is described by Eqn.\ref{Eq2}: 
$G_{xx}\propto 1/(P+1/P)$ is symmetric in $P$ 
with respect to $P = 1$ and 
the maximum appears at $P = 1$. On the basis of this 
symmetry we choose a point on each side of the 
$G_{xx}$ maximum. The associated values
of the backscattering function are 
$P_1 = W$ and $P_2 = 1/W$, respectively, with $W$ 
being a constant, except unity. We can write $P_1 = 
\exp(\Delta \nu_w/k)$ and $P_2 = \exp(-\Delta \nu_w/k)$ 
and obtain $P_1/P_2 = W^2 = \exp (2\Delta\nu_w/k)$, 
where $2\Delta \nu_w$ can be identified as the width of 
the $G_{xx}$ peak on the filling factor scale. This 
results in $\ln(W^2) = 2\Delta \nu_w/k = const$, which 
means that the temperature dependence of  $2\Delta \nu_w(T)$
and $k(T)$ must be the same, regardless any particular 
temperature dependence $k(T)$. 
The fact that all experimentally obtained traces of 
$R_{xx}^{ins}$  and  $R_{xx}^{top}$ of Ref.\cite{5}
collapse onto a single 
trace, if plotted with respect to $(\nu - \nu_c)T^{-\kappa}$,
suggests that the argument of the exponential 
function should have the form $\alpha (\nu - \nu_c)T^{-
\kappa}$ with $\alpha$ being a constant. 
However, as evident from above, also an 
alternative temperature dependence 
$k(T)=\alpha + \beta \cdot T$, 
which has been suggested
recently by Shahar et al \cite{Tlin}, can be used.

	In summary we have presented a model 
for the IQHE with a new representation of backscattering, 
which is shown to be in agreement with the general form of the 
Landauer-B\"uttiker formalism. It successfully  describes 
the transport regime also between IQHE plateaus. Even
though we use the edge channel approach for the
IQHE as an input for our model, the results are more general
and the model provides the correct framework to include
also dissipative bulk transport. Quite a number of well known 
facts can be obtained without needing 
the particular function 
of backscattering versus Landau level filling $P(\Delta \nu)$:  
(i) the semicircle relation between $\sigma_{xx}$ and 
$\sigma_{xy}$ for the complete QH regime as well as for 
the Hall insulator (HI) regime,  
 (ii) the critical value $\rho_{xx}^c=h/e^2$ in the 
HI regime,  
(iii) the value $\sigma_{xx}=\sigma_{xy}=0.5e^2/h$ at the 
critical point for a single Landau level,  
(iv) the maximum value $\rho_{xx}^{max}=h/4e^2$ for the 
1-2 transition,  
(v) the critical value $\rho_{xx}^c=h/5e^2$  for the 1-2 
transition, 
Using an exponential function for $P(\Delta \nu)$ 
we obtain 
further:  (vi) the validity of the relation $\rho_{xx}(\Delta 
\nu)=1/\rho_{xx}(-\Delta \nu)$ between the HI and the 
adjacent QH-liquid regime,
(vii) the equivalence of the temperature scaling of 
$\rho_{xx}$ in the HI regime, of $\rho_{xx}$ of the top LL 
and of the $\rho_{xx}$-peak width.  
(viii) regarding the temperature dependence
(using any $k(T)$ monotonously increasing with $T$), 
the model indicates correctly 
the existence of metallic like ($P<1$) and insulator 
like ($P>1$) regimes.

	Financial support came from the "Fonds zur F\"orderung der 
wissenschaftlichen Forschung", Austria (Proj.No. 
P10510NAW) and "Jubil\"aumsfonds der \"Osterreichischen 
Nationalbank", Austria (Proj.No. 6566). J.O. thanks D. Shahar 
for providing data in advance to publication.


\begin{references} 
\bibitem{Klitz} 	K. von Klitzing, G.Dorda, and 
M.Pepper, Phys. Rev. Lett. 45, 494 (1980)
\bibitem{6} S.A.Kivelson, D.H.Lee, and S.C.Zhang, 
Phys. Rev. B 46, 2223 (1992) 
\bibitem{1} M. B\"uttiker, Phys. Rev. B 38,  9375 (1988) 
\bibitem{2} D. Shahar, et al, Phys. Rev. Lett. 74, 4511 (1995) 
\bibitem{3} M.Shayegan, Solid State Commun. 102, 155 
(1997) 
\bibitem{4} D.Sahar, et al, Solid State 
Commun. 102 (12), 817 (1997) 
\bibitem{5} D.Shahar, et al,, Phys. Rev. Lett 79, 479 (1997)
\bibitem{7} I.Ruzin and Shechao Feng, 
Phys.Rev.Lett.74, 154 (1995) 
\bibitem{10}M.B\"uttiker, Semiconductors and Semimetals 
Vol. 35,  191 (1992),  Academic Press Inc. 
\bibitem{Kivel} J.K. Jain and S.A. Kivelson, 
Phys. Rev. Lett. 60, 1542 (1988); P.Streda, J.Kucera, A.H. 
MacDonald, Phys. Rev. Lett. 59, 1973 (1987) 
\bibitem{ec1} The EC-picture has to be understood as a 
formal treatment of transport, which refers to the potentials at 
the edge and where the sample current appears as a 
representative edge current. The usage of this theoretical 
concept must not be confused with giving an answer of the 
still remaining 
controversial question of how the carriers are indeed 
transported. 
\bibitem{8} J.Oswald, G.Span, Semicond. Sci.Technol. 
12, 345 (1997) 
\bibitem{9} The new aspect of this approach 
is an interpretation of EC-backscattering in terms of a 
dissipative bulk current which couples the edges. 
Alternatively to the Landauer-B\"uttiker formulation,
which is based on reflection and transmission 
coefficients, we obtain the 
longitudinal voltage drop from an 
edge current balance at the edges \cite{8}. 
\bibitem{11}In what follows we present an analytical 
version of our model. The results are fully in agreement 
with those of a numerical model which does not use the 
tensor relations. In the numerical model the sample 
current is allowed to flow via the two parallel EC systems, 
which are connected at the metallic contacts only. The 
potential differences at the contacts are then obtained in 
an iterative way from current conservation considerations. 
The numerical version of our model is able to give correct 
results for non-local contact configurations as well (considered 
for publication elsewhere). 
\bibitem{12}For a recent review see: S.L.Sondhi, 
S.M.Girvin, J.P.Carini, D.Shahar, Rev.Mod.Phys. 69, 315 
(1997);   
	see also B. Huckestein, Rev.Mod.Phys. 67,  357 
(1995) 
\bibitem{wende} In order to get a curve without a point of 
inflection at $\Delta\nu=0$, like experimentally observed,  
$\Delta\nu$ must appear linearly in the exponent. 
\bibitem{13}J.T.Chalker, P.D.Coddington, J.Phys.C 21, 
2665 (1988) 
\bibitem{14}Y. Huo, R.E. Hetzel, R.N. Bhatt, Phys. Rev. Lett 
70, 481 (1993) 
\bibitem{Tlin} D. Shahar et al., preprint cond-mat/9706045
\end{references}
\end{document}